\documentclass[]{emulateapj}

\usepackage{graphicx}
\usepackage{rotating}
\usepackage{pdfpages}

\slugcomment{}

\shorttitle{}
\shortauthors{Luchsinger et al.}

\begin{document}


\title{The host galaxies of micro-Jansky radio sources}


\author{
K.M.\ Luchsinger\altaffilmark{1,2}, 
M.\ Lacy \altaffilmark{2},
K.M.\ Jones\altaffilmark{3},  
J.C.\ Mauduit\altaffilmark{4},
J.\ Pforr\altaffilmark{5},
J.A.\ Surace\altaffilmark{4},
M.\ Vaccari\altaffilmark{6,7},
D.\ Farrah\altaffilmark{8},
E. Gonzales-Solares\altaffilmark{9},
M.J.\ Jarvis\altaffilmark{6,10}
C.\ Maraston\altaffilmark{11},
L.\ Marchetti\altaffilmark{12},
S.\ Oliver\altaffilmark{13}
J.\ Afonso\altaffilmark{14,15}
D.\ Cappozi\altaffilmark{11}
A.\ Sajina\altaffilmark{16}
\altaffiltext{1}{St Johns College, 60 College Avenue, Annapolis, MD 21401}
\altaffiltext{2}{National Radio Astronomy Observatory, 520 Edgemont Road, Charlottesville, 
VA 22903}
\altaffiltext{3}{Department of Astronomy, University of Virginia, 530 McCormick
Road, Charlottesville, VA 22904}
\altaffiltext{4}{Spitzer Science Center, California Institute 
of Technology, Pasadena, CA 91125}
\altaffiltext{5}{Laboratoire d’Astrophysique de Marseille, 
38, rue Fr\'{e}d\'{e}ric Joliot-Curie, 13388 Marseille cedex 13, France}
\altaffiltext{6}{Physics Department, University of the Western Cape, Private Bag X17, 7535, Bellville, Cape Town, South Africa}
\altaffiltext{7}{INAF Instituto di Radioastronomia, via Gobetti 101, 40129, Bologna, Italy}
\altaffiltext{8}{Department of Physics, Virginia Tech, Blacksburg, VA 24061}
\altaffiltext{9}{Institute of Astronomy, Madingley Rd, Cambridge CB3 0HA, UK}
\altaffiltext{10}{Astrophysics, Department of Physics, Keble Road, Oxford OX1 3RH, UK}
\altaffiltext{11}{Institute of Cosmology and Gravitation, Dennis Sciama Building, Burnaby Road, Portsmouth PO1 3FX, UK}
\altaffiltext{12}{Department of Physical Sciences, The Open University, Milton Keynes, MK7 6AA, UK}
\altaffiltext{13}{Astronomy Centre, Department of Physics and Astronomy, University of Sussex, Brighton, BN1 9QH}
\altaffiltext{14}{Observatorio Astronomico de Lisboa, Observatorio Astronomico de Lisboa, Portugal}
\altaffiltext{15}{Instituto de Astrof\'{i}sica e Ci\^{e}ncias do Espa\c co, Universidade de Lisboa, OAL, Tapada da Ajuda, PT1349-018 Lisboa, Portugal}
\altaffiltext{16}{Departamento de F\'{i}sica, Faculdade de Ci\^{e}ncias, Universidade de Lisboa, Edif\'{i}cio C8, Campo Grande, PT1749-016 Lisbon, Portugal}
\altaffiltext{17}{Department of Physics and Astronomy, Tufts University, 212 College Avenue, Medford, MA 02155}
}

\begin{abstract}
We combine a deep 0.5~deg$^2$, 1.4~GHz deep radio survey in the Lockman Hole with infrared
and optical data in the same field, including the SERVS and UKIDSS near-infrared
surveys, to make the largest study to date of the host galaxies of radio 
sources with typical radio flux densities $\sim 50 \;\mu$Jy. 87\% (1274/1467) 
of radio sources have identifications in SERVS to $AB\approx 23.1$ at 3.6 or 4.5$\mu$m, 
and 9\% are blended with bright objects (mostly stars), leaving only 4\% (59 objects)
which are too faint to confidently identify in the near-infrared.
We are able to estimate photometric redshifts for 68\% of the
radio sources. We use mid-infrared diagnostics to show that the source population consists of 
a mixture of star forming galaxies, rapidly accreting (cold mode) 
AGN and low accretion rate, hot mode AGN, with neither AGN nor starforming galaxies clearly 
dominating. We see the breakdown in the $K-z$ relation in faint radio source 
samples, and show that it is due to radio source 
populations becoming dominated by sources with radio luminosities 
$\sim 10^{23}\;{\rm WHz^{-1}}$. At these luminosities, both the 
star forming galaxies and the cold mode AGN have hosts with stellar 
luminosities about a factor of two lower than those of hot mode AGN, which 
continue to reside in only the most massive hosts. 
We show that out to at least $z\sim 2$, 
galaxies with stellar masses $>10^{11.5}\; M_{\odot}$ have a radio-loud 
fraction up to $\sim 30$\%. This is consistent with there being a sufficient number of 
radio sources that radio-mode feedback could play a role in 
galaxy evolution. 

\end{abstract}



\section{Introduction}

The radio source population at 1.4~GHz flux densities $S_{1.4}<1$~mJy contains a 
mixture of active galactic nuclei (AGNs), both radio-loud and radio-quiet, and 
star forming galaxies (Condon 1992, Afonso et al.\ 2005, Simpson et al.\ 2006). 
Radio-loud AGNs have, 
historically, been easier to study, being three to four 
orders of magnitude more luminous in the radio 
than starforming galaxies and dominating radio surveys at bright 
flux density levels. However, recent studies 
have shown that within populations with characteristic flux densities
of tens to hundreds of microJanskys (hereafter the $\mu$Jy population), 
the predominance of radio-loud AGNs gives 
way and the population begins to include significant numbers of 
radio-quiet AGNs and starforming galaxies 
(Jarvis \& Rawlings 2004; 
Simpson et al. 2006; Huynh et al. 2008; Smol\v{c}i\'{c} et al. 2008; 
Seymour et al. 2008; Strazzullo et al.\ 2010; Mao et al.\ 2012; 
White et al. 2012; Bonzini et al.\ 2013). Characterizing this mixed population
is important for using radio surveys to study the star formation history
of the Universe (Karim et al.\ 2011; Zwart et a.\ 2014), the cosmic history of AGN activity (Smol\v{c}i\'{c} et al.\ 2015),
and in order to use
radio continuum surveys for cosmological purposes.

Multiwavelength observations of radio sources 
allow the study of the relative distributions and properties of the three 
dominant types of sources (radio-loud AGN, radio-quiet AGN, and starforming galaxies) among the 
various wavelengths. Previous studies have used this multiwavelength technique to examine 
radio source populations in several deep fields, but have reached differing results as to the 
contributions of the different galaxy types. Most studies suggest a significant fraction 
(Huynh et al.\ 2008; Simpson et al.\ 2006; Smol\v{c}i\'{c} et al.\ 2008) or a majority 
(White et al.\ 2012; Seymour et al.\ 2008) of starforming and radio-quiet galaxies within the 
$\mu$Jy population, although some authors argue a continuation of the dominance of 
radio-loud AGNs from the Jansky and milliJansky populations to sub-milliJansky 
levels (Mignano et al.\ 2008). 

Deep radio surveys, combined with multiwavelength data, are essential for 
studying the nature of the host galaxies of the
AGN-powered radio source population, and,
as they are deep enough to contain a mixture of both 
radio-loud and radio-quiet objects, are helpful in our understanding of the origin
of radio-loudness in AGN. At high 1.4~GHz radio luminosities 
($L_{\rm 1.4}>10^{24}\; {\rm WHz^{-1}}$) radio sources have long been known to be hosted by the 
most massive galaxies at all redshifts. Early work by Lilly \& Longair (1984) and Lilly (1989)
showed that the near-infrared $K$-band magnitudes of radio galaxies formed a tight Hubble 
diagram. This so-called ``$K-z$ relation'' has been found to be a very useful tool for 
estimating the redshifts of distant, luminous radio sources, and is only very weakly 
dependent on radio luminosity for radio sources with $L_{1.4}\stackrel{>}{_{\sim}} 10^{24}\; {\rm Wm^{-2}}$ (Willott et al.\ 2003; McLure et al.\ 2004). The scatter 
in the $K-z$ relation towards fainter $K$-magnitudes has long been predicted to increase as 
fainter radio samples are studied, however (e.g.\ Jarvis et al.\ 2001), when radio source 
populations become dominated by progressively lower radio-luminosity AGN and starforming
galaxies. The first signs of this breakdown in the $K-z$ relation 
start to appear in $\mu$-Jy radio surveys (Simpson et al.\ 2012), which we 
confirm with a larger sample of sources in this paper.

Understanding the mix of 
galaxies that host the $\mu$Jy radio source population is essential for planning for 
future radio observatories such as the SKA, in 
particular for cosmological studies where different types of host galaxies cluster differently
(e.g.\ Ferramacho et al.\ 2014). We therefore decided to use some of the 
best currently-available multiwavelength data to investigate the nature of the
radio source host population. We base our multiwavelength catalog on 
the Spitzer Extragalactic Representative Volume Survey (SERVS), which uses deep Spitzer data to 
detect galaxies out to $z\sim 5$ (Mauduit et al.\ 2012). All five fields 
chosen by this survey were selected specifically to be integrated into multi-wavelength 
observations. In particular, the Lockman Hole field has two deep radio surveys within it, the very deep survey of Owen et al.\ (2009) in the north of the field and the wider, but somewhat shallower survey of Ibar et al.\ (2009) in the east. In this paper, we match the SERVS data with the Ibar et al.\ radio survey, and other surveys in the Lockman Hole field at optical through infrared wavelengths to investigate the host galaxy properties of the $\mu$Jy radio source population.
These data are summarized in Section 2. Details of the catalogs 
and band-matching are given in Section 3, and Section 4 discusses 
the quality of the photometric redshifts. The methodology for discriminating 
between AGN and starforming galaxies is presented in Section 4. Section 6 presents the 
properties of the radio sources, and conclusions may be found in  Section 7. 
A cosmological model 
of $\Omega_M = 0.3, \Omega_{\Lambda} = 0.7$, and $H_0 = 70 \; {\rm kms^{-1} Mpc^{-1}}$ 
is assumed throughout the text.


\section{Multiwavelength Observations}
\begin{figure}
  \plotone{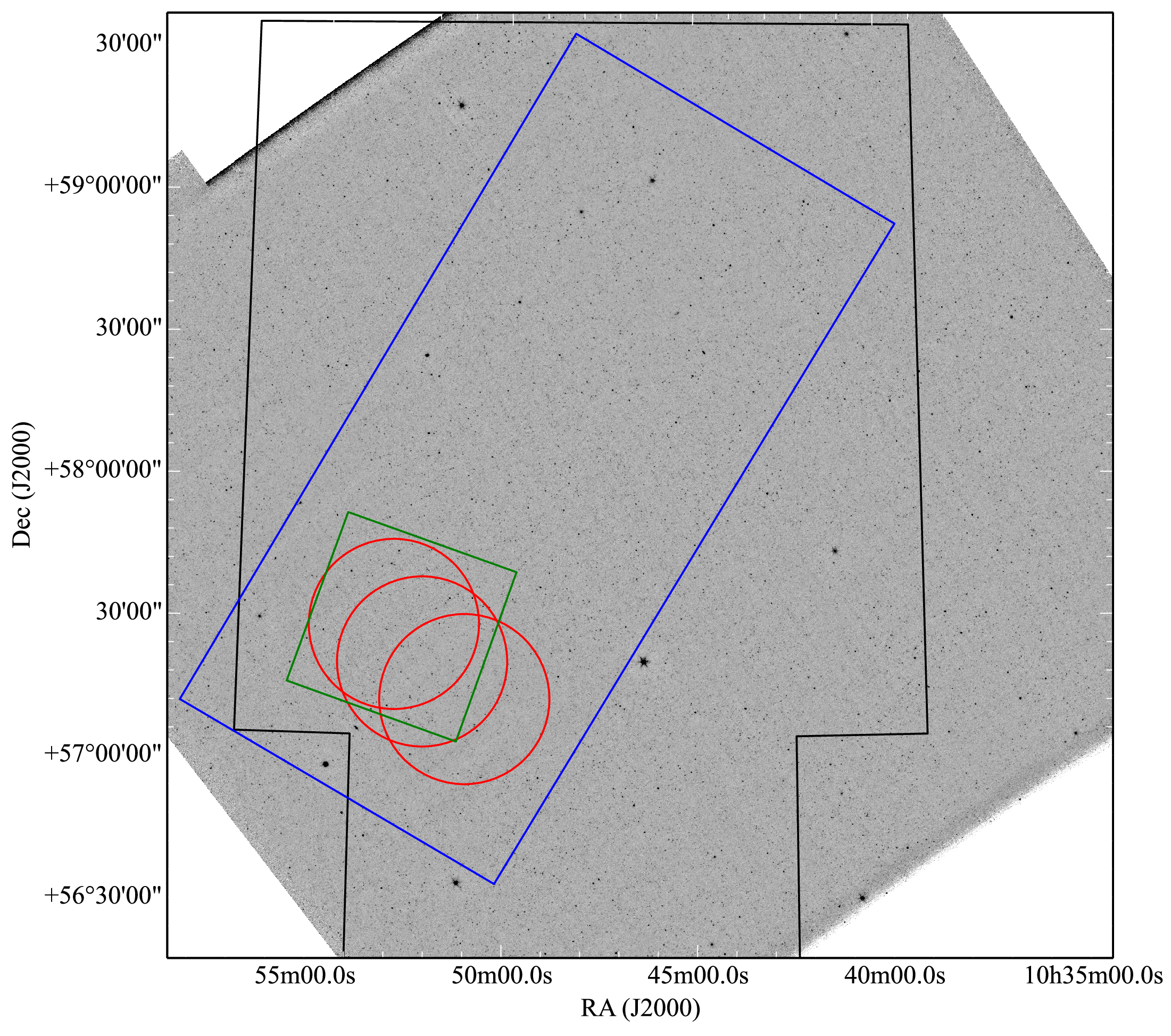}
  \caption{Surveys in the Lockman Hole used in this work. The greyscale shows the SWIRE 24$\mu$m image, which is close to coincident with both the IRAC SWIRE coverage, and the optical imaging survey of Gonz\'{a}les-Solares et al.\ (2011). The black outline shows the coverage of the UKIDSS DXS survey, the blue the SERVS coverage and the red circles the positions of the VLA pointings used for the Ibar et al.\ (2009) radio survey. The green square shows the deep (``level 3'') Herschel HerMES coverage (the whole field is covered to a lower depth), see Section 5.5.}\label{fig:lockman}
\end{figure}

The Lockman Hole is a particularly appealing 
candidate for a multiwavelength study because there have been many observations made of 
it in a wide variety of wavelengths. These data can be combined to form a very thorough 
study encompassing objects with redshifts out as far as 
$z\approx 5$. We have assembled a multiwavength dataset, detailed below, with which 
to study the host galaxies of $\mu$Jy radio sources in this field. Figure 
  \ref{fig:lockman} shows the spatial overlap of the datasets described below.

\subsection{SERVS data}

SERVS uses deep Spitzer data taken during the postcryogenic Spitzer Space Telescope mission over the course of 1400 hours using the Infrared Array Camera (IRAC) at wavelengths of 3.6 and 4.5~$\mu$m, and detects objects out to $z\approx 5$. The survey covers a total of 18~deg$^2$ to a detection limit of $\approx 2\; \mu$Jy ($AB=23.1$) in five fields, including 4~deg$^{2}$ in the Lockman Hole. SERVS catalogs were produced using Sextractor (Bertin \& Arnouts 1996) as detailed in Mauduit et al.\ (2012).

\subsection{Optical data}
Gonz\'{a}lez-Solares et al.\ (2011) observed the Lockman Hole with the Isaac Newton Telescope Wide 
Field Camera (INT WFC) in the $u, g, r, i,$ and $z$-bands, with $AB$ magnitude limits of 23.9, 24.5, 
24.0, 23.3, and 22.0, respectively. Supplementary observations were taken with the Mosaic 1 
camera on the Mayall 4-m Telescope of the Kitt Peak National Observatory (KPNO) in the $g, r,$ 
and $i$-bands. Details of the source extraction are given in Gonz\'{a}lez-Solares et al.\ (2011).

\subsection{Ground-based near-infrared data}

The UKIRT Infrared Deep Sky Survey (UKIDSS) observed the Lockman Hole as part of its Deep 
Extragalactic Survey (DXS) using the Wide Field Camera (WFCAM) of the United Kingdom 
Infrared Telescope (UKIRT) at wavelengths 1.2 $\mu$m and  2.2 $\mu$m, 
corresponding to the $J$ and $K$ bands, respectively. The survey reaches AB magnitude depths of $J_{AB}=23.1$ and $K_{AB}=22.5$. Lawrence et al.\ (2007) describe
the survey and the production of the catalogs.

\subsection{SWIRE data}
The Spitzer Wide-area Infrared Extragalactic (SWIRE) survey observed the Lockman Hole 
during its cryogenic mission using the Infrared Array Camera (IRAC) and the Multiband 
Imaging Photometer for Spitzer (MIPS). IRAC observed the Lockman Hole using all four 
channels, 3.6~$\mu$m, 4.5~$\mu$m, 5.8~$\mu$m, and 8.0~$\mu$m; and MIPS observed the 
Lockman Hole with all three of its channels, 24~$\mu$m, 70~$\mu$m, and 160~$\mu$m.
The SWIRE catalogs reach depths of $\approx 10\; \mu$Jy ($AB \approx 21.4$) at 3.6 and
  4.5~$\mu$m. SWIRE detected galaxies out to $z\approx 3$ (Lonsdale et al. 2003). Details of the 
data release used may be found in the SWIRE data release 2 
document\footnote{http://irsa.ipac.caltech.edu/data/SPITZER/SWIRE/docs/delivery\_doc\_r2\_v2.pdf}.

\subsection{Radio data}
Part of the Lockman Hole was observed at frequencies of 610 MHz and 1.4 GHz 
using the Giant Metre-wave Radio Telescope (GMRT) and the Very Large Array (VLA), 
respectively by Ibar et al.\ (2009). Their survey covered $\approx 0.5$~deg$^2$
in the Southern region of the Lockman Hole (centered on RA, Dec 163.0$^{\circ}$,57.35$^{\circ}$, generally called the ``Lockman Hole East'') and contains 1452 radio sources detected at 1.4GHz, comprised of 1467 discrete components.
The GMRT data reach an RMS noise of $\approx 15\; \mu$Jy/beam, and the VLA data an RMS of 
$\approx 6\; \mu$Jy/beam in the center of the surveys, where the sensitivity is 
highest, tapering off to  $\approx 10\; \mu$Jy/beam at 1.4~GHz towards the edge of the
survey region (see Figure \ref{fig:fluxhist} for the flux density distribution).The depth of the GMRT data is such that nearly all of the 1.4GHz sources have counterparts at 610MHz.

\section{Catalog matching}
Data from the optical through far-infrared surveys were combined 
into a SERVS ``Data Fusion'' catalog\footnote{http://mattiavaccari.net/df}, 
using a matching radius of 1.0$^{''}$
to match them to the SERVS dual-band catalog.
This catalog contained 951102 objects covering the full 4~deg$^2$ of the
SERVS Lockman area. Details of this catalog will be presented in Vaccari et al.\ (in preparation).
This Fusion catalog was
then matched to the 1467 individual component detections at 1.4~GHz
(with peak-to-noise ratio $>5$) 
in the Ibar et al.\ (2009) catalog.
The radio sources were matched to the SERVS Fusion catalog using a 
selection radius of 1.2 arcseconds (compared to a 4 arcsecond beamsize), 
which resulted in 1245 matched objects - a 85\% 
match rate down to the $AB \approx 23.1$ catalog limit. 
We then compared the remaining 222 unmatched radio components with the SERVS
data manually, and matched an additional 29 radio components, six of which we believe are
components of multi-component radio sources, and 23 of which were matched
to faint SERVS sources below the catalog limit, 
resulting in a 87\% match rate. Of the remaining unmatched objects, 
134 (9\% ) were confused by brighter objects (mostly stars and bright galaxies) in SERVS.
Of the remainder, only 31 (2\%) appeared to be completely blank fields, and a further
28 were most likely blank, but were within $\approx 2{''}$ of faint objects in the field. These
statistics emphasise the rarity of infrared-faint radio sources (Norris et al.\ 2011).

To evaluate the number of matches expected at random, we offset the positions of the radio sources by approximately 1-arcmin in RA and Dec, and redid the matching four times with different offsets. The mean number of random matches was 90, corresponding to 6\% of the number of radio components, which we treat as negligible.

\section{Photometric redshifts}

\subsection{Determination of Photometric Redshifts}
Photometric redshifts of the objects observed in the SERVS study were 
obtained by Pforr et al. (in preparation) using the Hyper-z photometric redshift code
(Bonzella, Miralles \& Pello 2000).
A correlation of the photometric redshifts with the spectroscopic redshifts
of AGN in the field (see below) is described in the Appendix, and shows that 
$\sim 85$\% of the redshifts are accurate to within an RMS of 
$\Delta z/(1+z) \approx 0.06$, the remainder being outliers. Highly-luminous AGN
at high redshifts, where the hot dust from the AGN in the IRAC 3.6 and 
4.5$\mu$m bands can affect the redshift 
estimate, may have poorer photo-zs, however. 

\begin{figure*}
\plotone{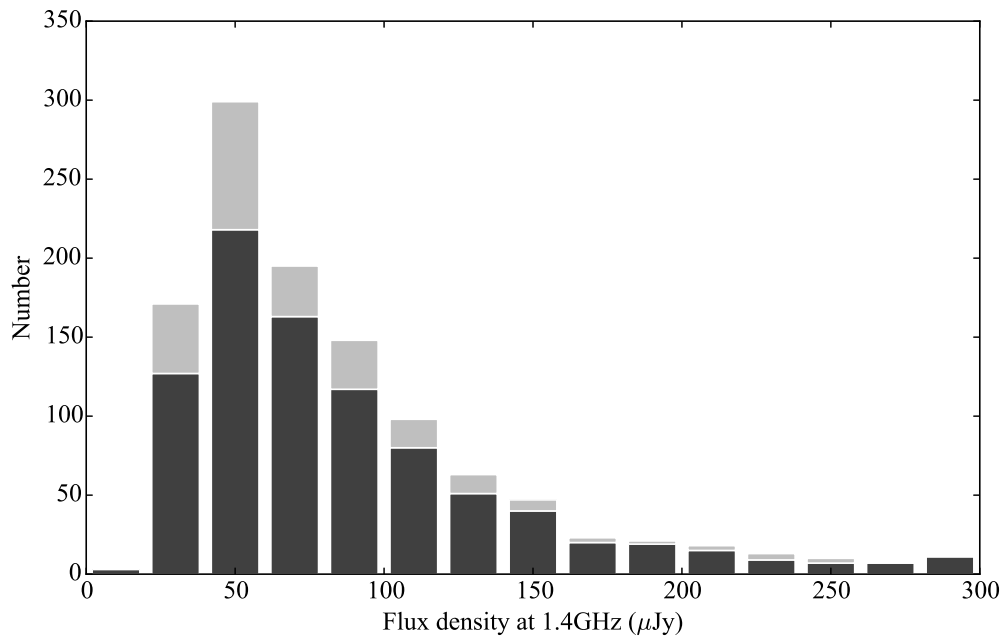}
\caption{Histogram of flux densities at 1.4~GHz for our sample. The dark grey indicates
objects with $\geq$ 4-band photometric redshifts, the lighter grey objects lacking 
photometric redshifts.}\label{fig:fluxhist}
\end{figure*}

Of the photometric redshifts for the radio galaxies, 20 (1\%) failed to find a solution. 
All of these failed solutions
had only limits on their detection in either or both of the $J$-band or $K$-band. To ensure a 
good quality photometric redshift, we thus excluded the 254 objects
(20\% of the 1245 objects matched between the radio and SERVS) lacking 
a $J$- or $K$-band detection from any analysis that needed redshift information, 
ensuring that we had a minimum of four bands to estimate a redshift ($J$, $K$, [3.6]
and [4.5]). In Figure \ref{fig:fluxhist} we show the 1.4~GHz flux density distribution 
of the survey for objects with and without photometric redshifts, indicating that 
there is not a very strong bias towards, for example, only the brighter radio sources
having photometric redshift information.
The photometric redshifts extend out to 
$z\approx 5$, as seen in Figure \ref{fig:zhist}
(though some of the redshifts at $z>4$ use only the minimum four bands, and so 
may be unreliable). There is 
a dip in the redshift distribution 
at $z \approx 1.4$ which has no clear origin. Two possible explanations are that
 it may be a degeneracy in the
photometric redshifts in the region of the ``redshift desert'' where the optical 
bands in particular 
do not contain strong spectral features to constrain the redshifts, or that it may 
reflect real large-scale structure in the field. Spectroscopic redshifts were available for 62 of the AGN from Lacy et al.\ (2013) or the SDSS quasar survey (Ahn et al.\ 2014), including the normal type-1 AGN whose optical emission is dominated by quasar light. We have used these in preference to the photometric redshifts where applicable.


\subsection{Predictions of Photometric Redshifts for $\mu$Jy radio sources}
We used the SKA Simulated Skies (S$^3$) 
extragalactic model (Wilman et al.\ 2008) that uses measured luminosity functions
combined with an underlying numerical dark matter simulation
to obtain a predicted redshift distribution for 
$\mu$Jy radio sources. This predicted redshift distribution is compared 
to our measured one in Figure \ref{fig:zhist}. Both distributions peak at $z\approx 0.7$, though
our measured distribution has fewer sources in the lowest redshift bin. 
This deficit is also seen in the smaller (but much more spectroscopically-complete) sample
of Simpson et al.\ (2012), so is unlikely to be due to large-scale structure fluctuations,
though we cannot rule out incompleteness at very low redshifts due to galaxies being 
resolved out in the radio survey.
We see fewer sources in the $z>2.5$ bins than the 
predictions, this may be explained
though by the 20\% of faint sources we excluded due to missing $J$- and/or $K$-band 
detections. The
median redshift we observe, 0.99, is very similar to the median redshift predicted
by the simulations (1.04).

\section{Object classification}

We used a combination of a mid-infrared color-color diagram and the radio to mid-infrared
flux ratio to classify as many of the radio sources as possible into three overall
classes, namely cold-mode AGN (high accretion rate objects with dynamically-cold 
accretion flows and Eddington Luminosities 
$L{_{Edd}}\sim 0.01-1$), hot-mode AGN (low accretion rate objects accreting gas at about the
galaxy virial temperature of $\sim 10^{6}$K, with 
$L_{\rm Edd}\stackrel{<}{_{\sim}}0.01$, and jets possibly powered by the spin of the black hole, 
e.g.\ Mart\'{i}nez-Sansigre \& Rawlings [2011]) and star-forming galaxies. 
Mao et al.\ (2012) show that mid-infrared classifications are
broadly consistent with those from spectrosopy.

\subsection{Rapidly accreting, cold-mode AGN}
When looking for the relative contributions of AGN and starforming galaxies, bright
AGN are simpler to isolate than starforming galaxies. For the 479 objects
that were detected in all four of the 3.6, 4.5, 5.8 and 8.0$\mu$m bands of the 
Infrared Array Camera (IRAC) in the SWIRE data, we were able to isolate 
rapidly-accreting, cold-mode AGN using the 
method proposed by Lacy et al.\ 
(2004, 2007), which selects AGN based on their color relative to starforming or quiescent
galaxies. Cold-mode AGN have a characteristic red power-law in the mid-infrared that stands out 
well in a ${\rm log_{10}}(S_{5.8}/S_{3.6})$ versus ${\rm log_{10}}(S_{8.0}/S_{4.5})$
color-color plot, (Figure \ref{fig:iccd}, left). We used $S_{3.6}$ and $S_{4.5}$ from 
SERVS and $S_{5.8}$ and $S_{8.0}$ from SWIRE. Three cuts were made: 
${\rm log_{10}}(S_{5.8}/S_{3.6}) > -0.1$, ${\rm log_{10}}(S_{8.0}/S_{4.5}) > -0.2$, 
and ${\rm log_{10}}(S_{8.0}/S_{4.5})\leq 0.8 {\rm log_{10}}(S_{5.8}/S_{3.6})$, which 
produce the wedge-shaped area in the upper right corner of Figure \ref{fig:iccd}. 
This area represents the objects likely to be cold-mode AGN. These 226 objects
make up 15\%  of the radio sources in the Ibar et al.\ survey (Figure \ref{fig:pie}). 
Spectroscopic follow-up of AGN samples selected in this manner show
that this particular ``wedge'' selection is reliable in the sense that 78\%
(527 out of 672) of
objects for which spectroscopic redshifts and
classifications could be obtained using optical or near-infrared spectra 
are confirmed as AGN (Lacy et al.\ 2013). (Many of the remainder may also be highly-obscured 
AGN, but cannot be confirmed as such from optical/near-infrared data alone.)
Among the cold mode AGN there are a 
total of six type-1 quasars from the Sloan Digital Sky Survey Data Release 
10 (Ahn et al.\ 2014) 
(five of which are in the Lacy et al.\ [2013] sample).

\begin{figure*}
\plotone{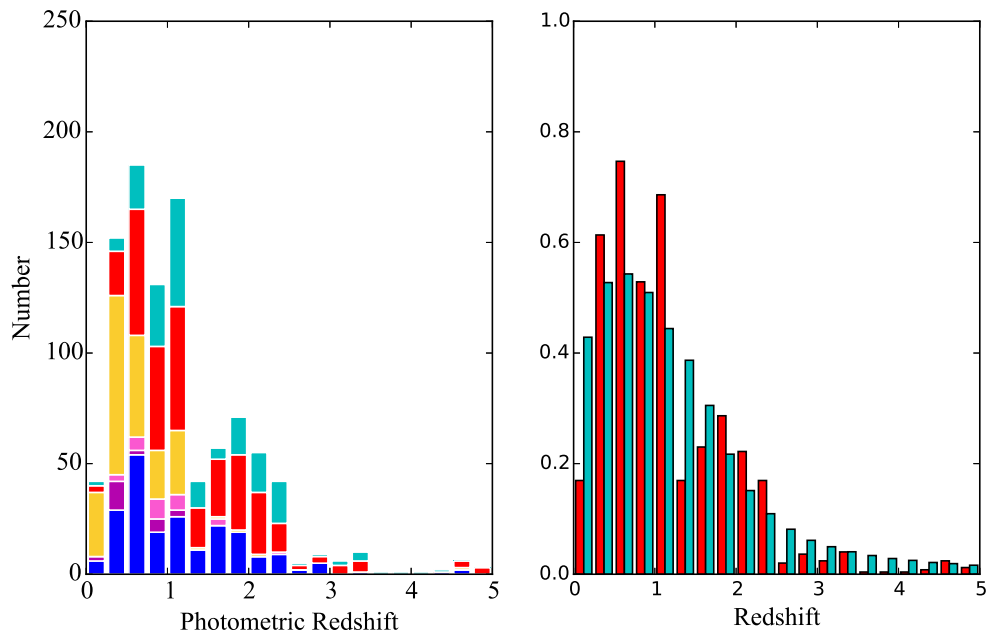}
\caption{Left: photometric redshift distribution by type (see Section 5). Blue indicates actively
  accreting (cold mode) AGN; dark magenta, hot-mode AGN (based on the IRAC color-color plot);
    light magenta, likely hot-mode AGN (based on $q_{24}$); red, ambiguous AGN/starforming 
objects, and cyan, unclassifiable objects.
Right: the red histogram shows the normalized photometric redshift distribution for our sample, 
the cyan one the predicted redshift distribution from the full volume of the 
SKA simulations (Wilman et al.\ 2008).}\label{fig:zhist}
\end{figure*}

\begin{figure*}
\plotone{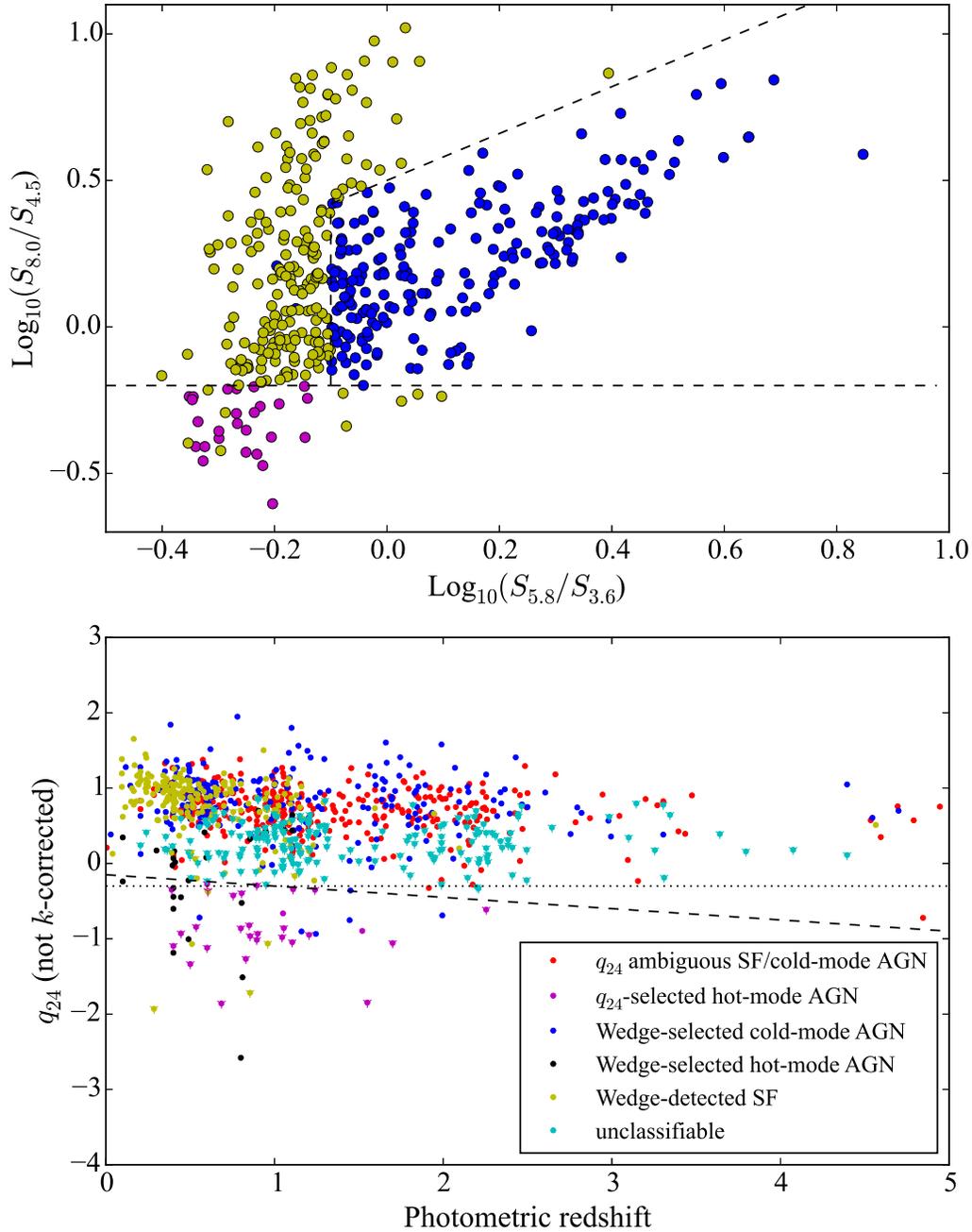}
\caption{Classification of radio sources. 
Left: infrared color-color diagram. The four lines separate the 
objects into three areas: 
the wedge-shaped area in the upper right corner contains likely cold-mode AGN (blue
dots); the 
area occupying the lowest two corners contains likely hot-mode AGN (black dots); 
and the area in the 
upper left corner contains likely starforming galaxies (yellow dots). Note
there are a few objects classified as starbursts in the hot-mode AGN 
region, these are Herschel detections and are mostly $z>1$ ULIRGs
(see Section 5.5). Right: 
$q_{24}={\rm log_{10}}(S_{\rm 24}/S_{\rm 1400})$ (not $k$-corrected) against redshift for the radio sources with matches
at 24$\mu$m in the SWIRE catalog. Downward pointing triangles 
indicate upper limits. The dashed line indicates the criteria we 
used to separate radio-loud hot-mode AGN from cold-mode AGN and 
starbursts for objects with photometric redshifts, the dotted line for those 
without (see text for details).}\label{fig:iccd}
\end{figure*}

\begin{figure*}
\plotone{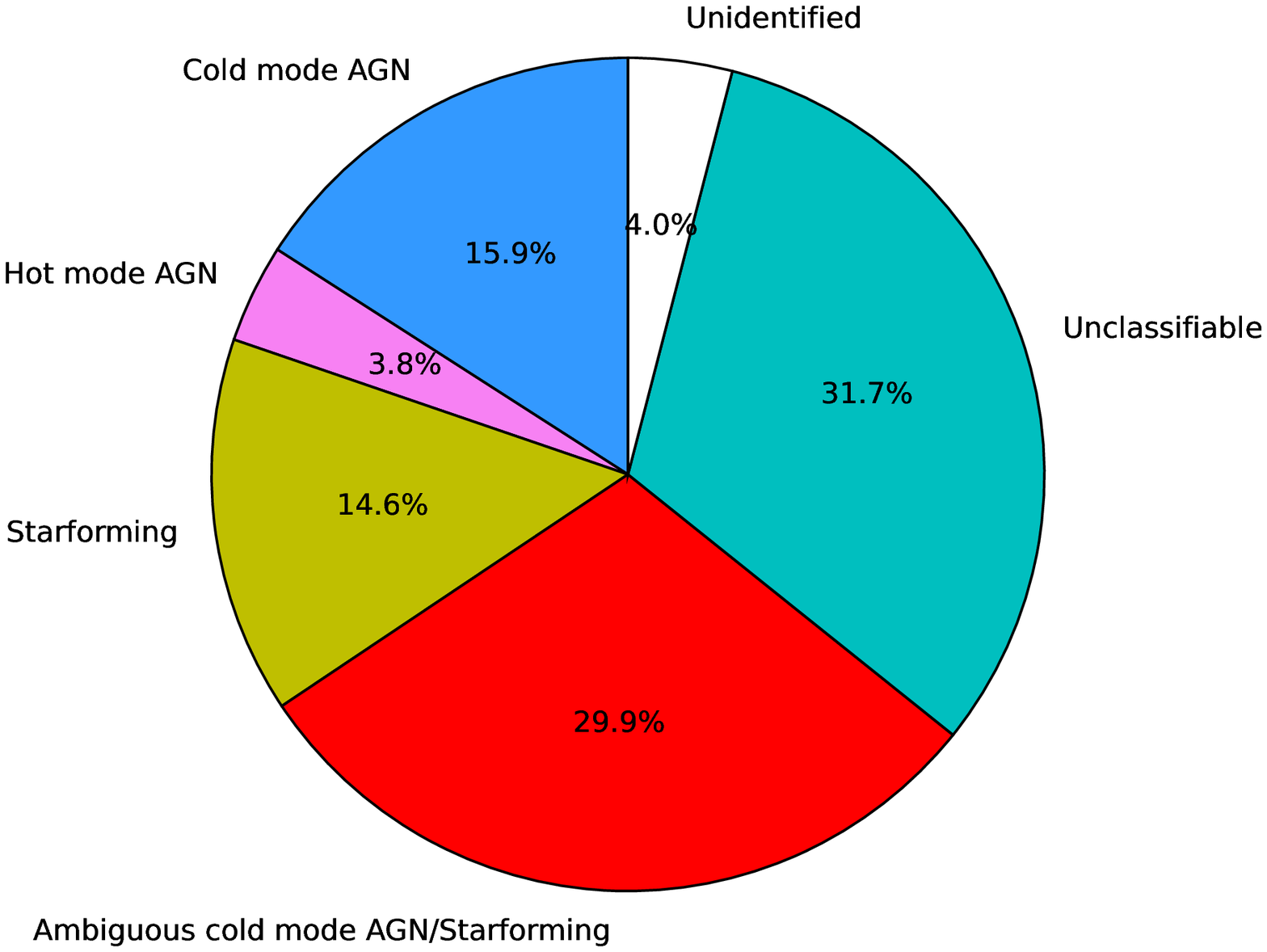}
\caption{Pie chart for 
the whole Ibar et al.\ 1.4~GHz survey using our 
classifications described in Section 5. The numbers of objects in each class are as follows:
  226 cold mode AGN, including six type-1 quasars; 74 hot mode AGN (comprising of 37 from the 4-band IRAC classification and 37 from the $q_{24}$ classification); 207 star-forming galaxies; 437 objects ambiguous between star formation and cold-mode AGN; 464 unclassifiable objects (excluding those that are definitely unidentified in SERVS), and 59 unidentified objects.}
\label{fig:pie}
\end{figure*}

\subsection{Low accretion rate, hot-mode radio AGN}
Lacy et al.\ (2006), Afonso et al.\ (2011), Gurkan et al.\ (2014), and Singh et al.\ (2014) 
also find that some weak radio AGN appear outside the AGN or starforming galaxy selection 
region, with IRAC colors similar to quiescent galaxies. 
These galaxies are likely the low accretion rate, ``hot mode'' AGN, whose 
radio jets are highly efficient.  
These radio sources can be found in the area beneath the cut formed by 
the line ${\rm log_{10}}(S_{8.0}/S_{4.5}) = -0.2$ in the left-hand panel of 
Figure \ref{fig:iccd}.

We also use the $q_{24}={\rm log_{10}}(S_{24}/S_{1400})$
value to classify objects without 4-band IRAC detections
(e.g.\ Ibar et al.\ 2008). 
As the SWIRE 24$\mu$m flux 
density limit, $\approx 150 \mu$Jy, is five times $S_{1.4 \; {\rm GHz}}$ of 
our faintest radio sources, and given that the typical value of $q_{24}$ for star forming 
galaxies is $\approx 1$  (Appleton et al.\ 2004; Ibar et al.\ 2008), 
we expect to detect the majority of star-forming galaxies and cold-mode AGN 
at 24~$\mu$m. We indeed find that the fraction
  of radio sources detected at 24$\mu$m is high (overall,
  874/1245 (70\%) of matched
  radio sources are detected at 24~$\mu$m).
  Thus, the radio-bright objects not detected at 24$\mu$m
  are likely to be hot-mode AGN.
We place the dividing line between 
radio-loud and radio-quiet at $q_{24}\approx 0$ at $z=0$, then use the template $k$-corrections
shown in figure 1 of Ibar et al.\ (2008) to approximate the change in redshift of this 
dividing line as $q_{24}<-0.15(1+z)$ (the dashed line in Figure 1, right-hand panel). 
For those objects not detected at 24~$\mu$m
we assume a limit of 150~$\mu$Jy to classify them. We separate the hot mode AGN 
selected using the wedge-based and
$q_{24}$-based methods in Figures \ref{fig:zhist} and \ref{fig:iccd}, 
but thereafter include them all in a single class, that comprise about 4\% of the 
total radio source population in this sample.

For the objects lacking photometric redshifts we approximated a classification by assuming
a mean redshift of unity (dotted line in the right-hand panel of Figure \ref{fig:iccd}). As
can be seen, this should result in $\stackrel{<}{_{\sim}}10$ misclassifications. 

\subsection{Star-forming galaxies and 24~$\mu$m detected objects with ambiguous classifications}
The remainder of the objects in the infrared color-color diagram -- the upper left 
corner in Figure \ref{fig:iccd} - correspond to starforming galaxies, 
and make up 15\% of the radio sources. 
In addition, 30\% of objects are detected at 24$\mu$m with $q_{24}$ above
the dashed (or dotted, in the case of objects lacking photometric redshifts) 
line in the right-hand panel of Figure \ref{fig:iccd}, but not 
detected in all four
IRAC bands. These most likely comprise a mix of cold-mode radio AGN 
and star-forming galaxies, and we henceforth classify them
as ``ambiguous AGN/SF''. 

\subsection{Unclassifiable objects}

32\% percent of the identified radio sources remain unclassifiable using either the 
infrared color-color diagram or $q_{24}$, because they lack 4-band IRAC detections, 
are undetected at 24~$\mu$m and their
flux density limit at 24~$\mu$m is insufficiently deep 
compared to their radio fluxes to classify them as radio-loud or radio-quiet. 


\subsection{Comparison of wedge and $q_{24}$ selection}

For the most part our classification scheme is self-consistent between the wedge technique
and the $q_{24}$ one. Amongst the objects with spectroscopic redshifts, however, 15/37 of the
hot mode AGN that are classified as such by the wedge technique show up as radio-quiet in the
$q_{24}$ plot. Furthermore, in eight cases, objects initially classified as hot-mode AGN
are detected in Herschel 
HerMES (Oliver et al.\ 2012) images at 250~$\mu$m and also have photometric redshifts $>1$. 
These are most likely 
mis-classified star-forming Ultraluminous Infrared Galaxies (ULIRGs), 
which, due to their high redshifts, no longer have the PAH 
features in the IRAC 8$\mu$m band resulting in them falling out of the star formation
region in Figure \ref{fig:iccd} (left). They have been reclassified as star-forming galaxies.
There is also one low redshift ($z=0.04$) galaxy which appears to be a HerMES
detection, we also reclassify it as 
a starburst. The remainder tend to plot in the radio-louder side of the infrared-radio
correlation. They may have significant residual star formation, or 24$\mu$m emission from their
AGN (although hot-mode radio AGN tend to be faint in the mid-infrared, some do have significant
mid-infrared emission, e.g.\ Ogle, Whysong \& Antonucci [2005]). 
Similarly, there are 4/207 galaxies classified as star forming in the wedge selection
that appear radio-loud in the $q_{24}$ one, these may be radio-loud AGN in starburst 
galaxies. Further, more accurate, classification could be attempted
with full SED fitting, including Herschel data from the HerMES
survey (Oliver et al.\ 2012), but we will defer this activity to 
a future paper.

Our classification results can be compared to the study of Bonzini et al.\ (2013), who used a survey of similar depth in the Extended Chandra Deep Field South to classify a sample of 883 radio sources as 19\% radio-loud AGN, 24\% radio-quiet AGN and 57\% starforming galaxies. The main difference between our classification techniques and those used by Bonzini et al. is their use of a less conservative $q_{24}$ criterion for defining radio-loud AGN (called ``hot mode'' AGN in this paper), and also their X-ray data and deeper {\em Spitzer} data that allows them to resolve more ambiguities between cold-mode AGN and starforming galaxies.

\section{Properties of radio sources}

\subsection{Radio luminosity and spectral index}
\begin{figure*}
\plotone{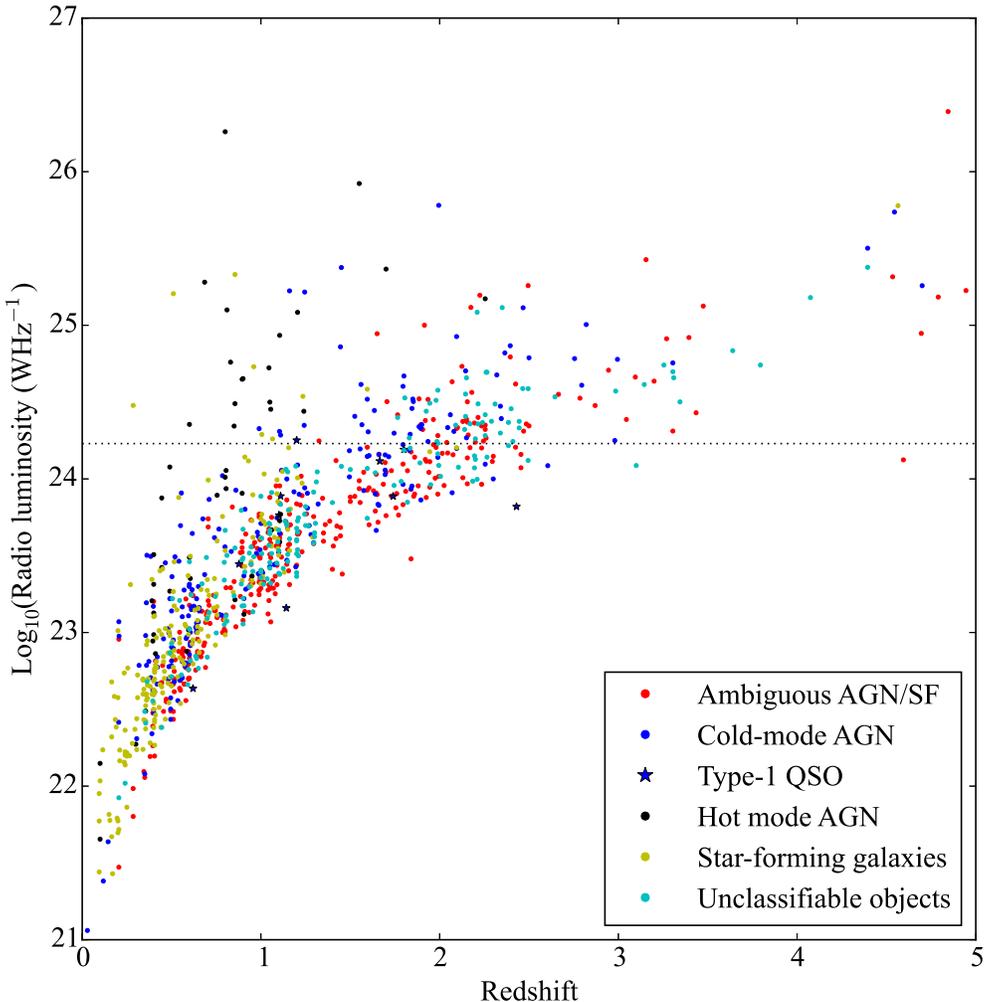}
\caption{Radio luminosity versus redshift for the radio sources, classified according to
their mid-infrared properties. The dotted line indicates a radio 
luminosity from star formation of 1000$M_{\odot} \; {\rm yr^{-1}}$, close to the
highest seen in starbursts (e.g.\ Karim et al. 2013) assuming a Salpeter
IMF, objects with ambiguous AGN/SF classifications (red dots) above this line are most likely to be AGN-dominated. As discussed in Section 4.1, the majority of our redshifts are photometric, however we include 62 spectroscopic redshifts.}
\label{fig:lumz}
\end{figure*}

In Figure \ref{fig:lumz}, we plot the radio luminosity versus the photometric 
redshifts of the radio 
sources. The radio luminosity was calculated for objects within 
the $\mu$Jy population using the flux densities and spectral indices of the 1.4 GHz objects 
within the Ibar et al.\ (2009) survey. Most of our radio sources have 
$L_{1.4}\sim 10^{22.5} - 10^{24.5}\; {\rm WHz^{-1}}$, spanning the
traditional boundary in luminosity between radio-loud and radio-quiet objects.
With few exceptions (probably mis-classifications) all the 
star-forming galaxies selected via Figure 1
have $L_{1.4}<10^{24.5}\; {\rm WHz^{-1}}$, corresponding to star formation 
rates $\stackrel{<}{_{\sim}}2000 \; M_{\odot}{\rm yr^{-1}}$, assuming a Salpeter IMF.

The Ibar et al.\ catalog includes spectral indices between 610MHz and 1400MHz
($\alpha^{610}_{1400}$) for objects detected at both frequencies. The mean
$\alpha^{610}_{1400}$ for the objects matched to SERVS is -0.78. The mean
spectral indices for the cold-mode AGN and star forming galaxies are very
similar to the overall mean (-0.77 and -0.79, respectively). That for the hot-mode AGN is significantly flatter, -0.44, indicating a large fraction of
flat-spectrum AGN in that class.

\subsection{K-band emission}
\begin{figure*}
\plotone{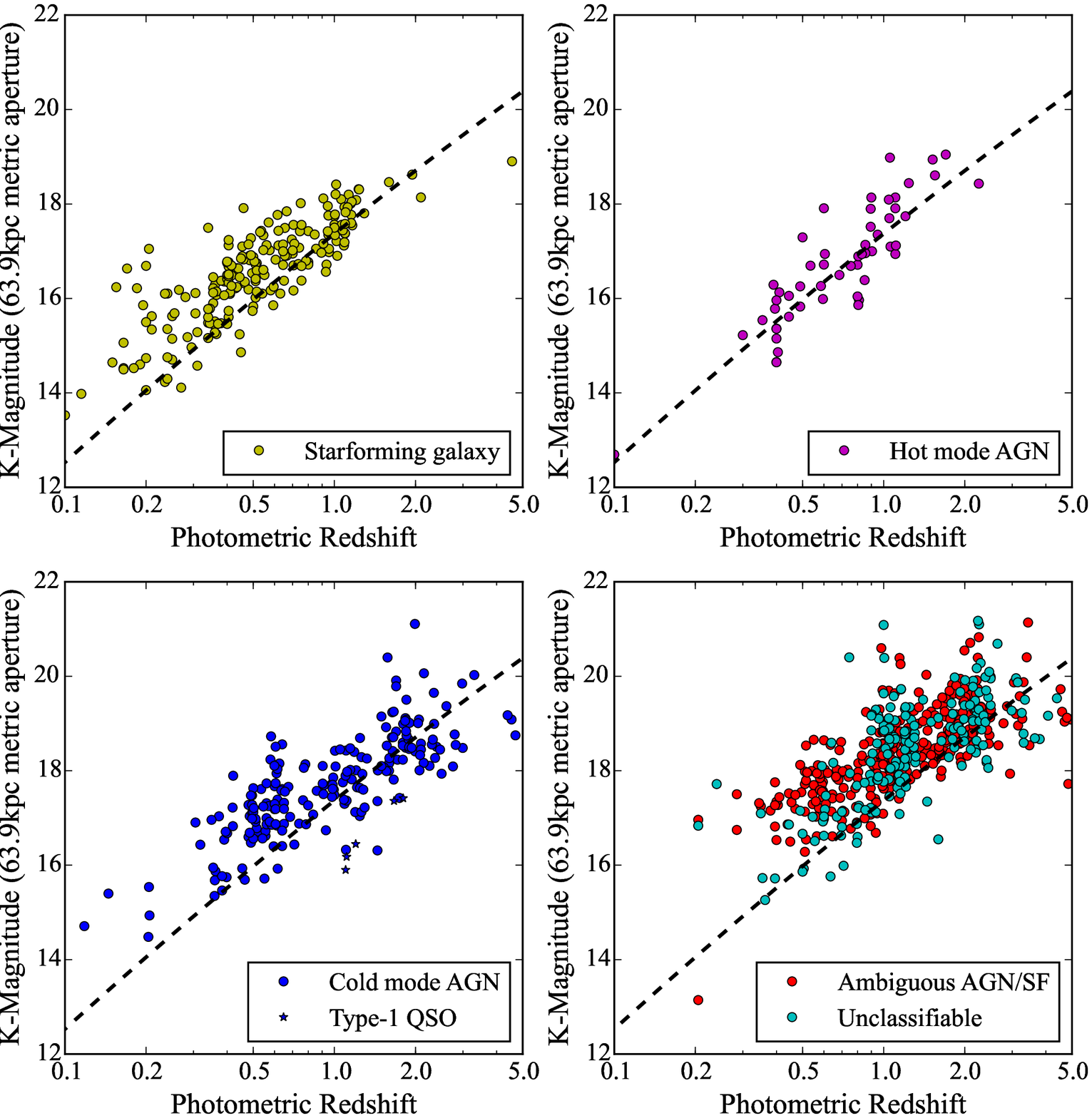}
\caption{The $K-z$ relation, split by radio source type as defined in Section 5. The dashed line
indicates the $K-z$ relation for radio sources from much brighter surveys as 
defined by Willott et al.\ (2003). The
hot-mode AGN are the only class of object that appears to trace the same $K-z$ relation
as samples of radio sources selected 
from surveys $\sim 1000$ times brighter in radio flux density.}
\label{fig:kz}
\end{figure*}

$K$-band emission from radio galaxies tends to consist primarily of stellar light 
(Lacy et al.\ 1995; Best, Longair \& R\"{o}ttgering 1998; 
Simpson, Rawlings \& Lacy 1999; Seymour et al.\ 2007), 
although in very luminous cold-mode AGN there
can be significant contributions from reddened quasar continuum
(Rawlings et al.\ 1995) or emission lines (Eales \& Rawlings 1993). Previous results 
generally found that the $K-z$ relation 
corresponded to a curve representing a stellar population which formed its stars at very high 
redshifts ($z > 5$) and passively evolved afterwards (e.g.\ Lacy, Bunker \& Ridgway 2000;
Jarvis et al.\ 2001; Willott et al.\ 2003). 
In Figure \ref{fig:kz} we plot the $K$-magnitudes against the 
logarithms of their photometric redshifts. 
The dashed curve in Figure \ref{fig:kz}, whose function is $ K = 17.37+4.53\;{\rm log_{10}} (z) - 0.31\;({\rm log_{10}} (z))^2$ , was proposed by Willott et al.\ (2003) as the polynomial 
of best fit for luminous steep-spectrum radio galaxies within a standard 63.9kpc 
metric aperture. 
Willott et al.\ show that their $K-z$ relation
fit closest to a model in which most stars formed in a single burst
at redshifts of $z = 10$, and evolved passively afterwards. However, 
the scatter towards brighter magnitudes at $z>2$ apparent in both Willott et al.\
and in Figure \ref{fig:kz} cannot eliminate a model in which many radio source hosts
formed at a lower redshift, $z\sim 5$. This is discussed further in Section 6.3 below.

\begin{figure*}
\plotone{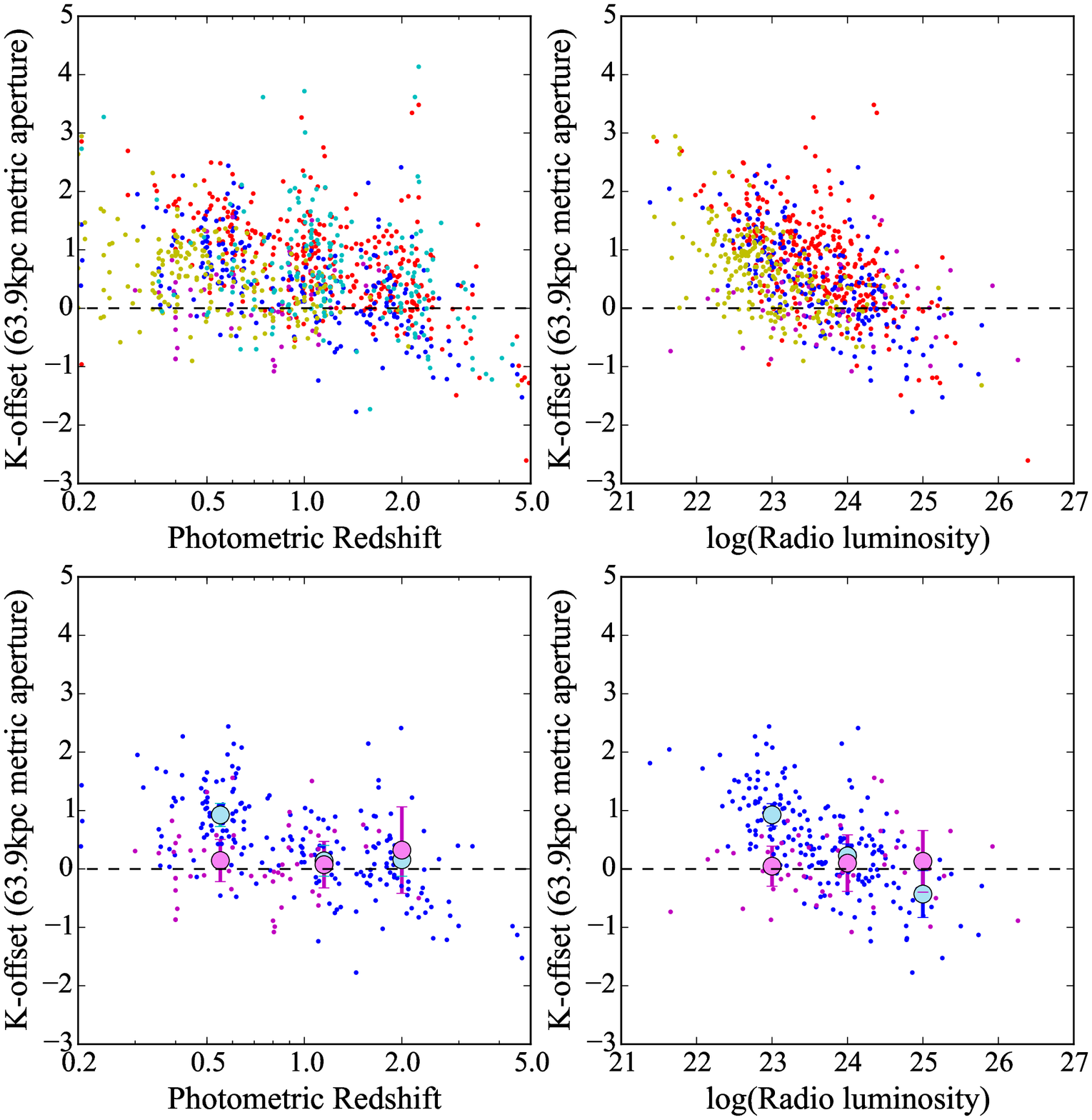}
\caption{Offsets from the $K-z$ relation of Willott et al.\ (2003) as a function of
redshift and radio luminosity (symbol colors as in Figure 4). Top panels, all radio source types. Bottom panels, AGN-powered
radio sources only. The large grey and cyan points represent averages, with 3~$\sigma$ 
error bars, of the hot-mode and cold-mode AGN, respectively.}
\label{fig:kzoff}
\end{figure*}

Figure \ref{fig:kz} shows that most of the hosts of $\mu$Jy radio sources 
are fainter in $K$-band than the mean for
the hosts of the much radio-brighter sources from the 3C, 6C and 7C samples
(with typical radio flux densities at 1.4~GHz $\sim 50-1000$~mJy) studied by 
Willott et al.\ (2003). This is perhaps
not surprising for the starforming galaxy population, which has a very different
physical mechanism for radio emission than the AGN populations that dominate the
Willott et al.\ sample, but it also seems
true for the cold-mode AGN population, at least at low redshifts/luminosities. 
In Figure \ref{fig:kzoff} 
we plot the offset from the $K-z$ relation
for different types of object (especially AGN) as a function of radio luminosity and redshift. In the
lowest radio luminosity bin, $10^{22.5-23.5}\; {\rm WHz^{-1}}$, the mean and error in the 
mean of the offset is $0.93\pm 0.06$ for cold-mode AGN (significant at 15$\sigma$), 
but insignificant ($0.05\pm 0.11$) for hot-mode AGN. In the next bin, 
$10^{23.5-24.5}\; {\rm WHz^{-1}}$, the offsets is barely significant (0.22$\pm 0.07$ for
the cold-mode AGN and again insignficant 0.10$\pm 0.16$ for the hot-mode AGN), and similarly
we see no significant offsets in the 
highest luminosity bin ($10^{23.5-24.5}\; {\rm WHz^{-1}}$). 

\begin{figure}
\includegraphics[angle=90,scale=0.35]{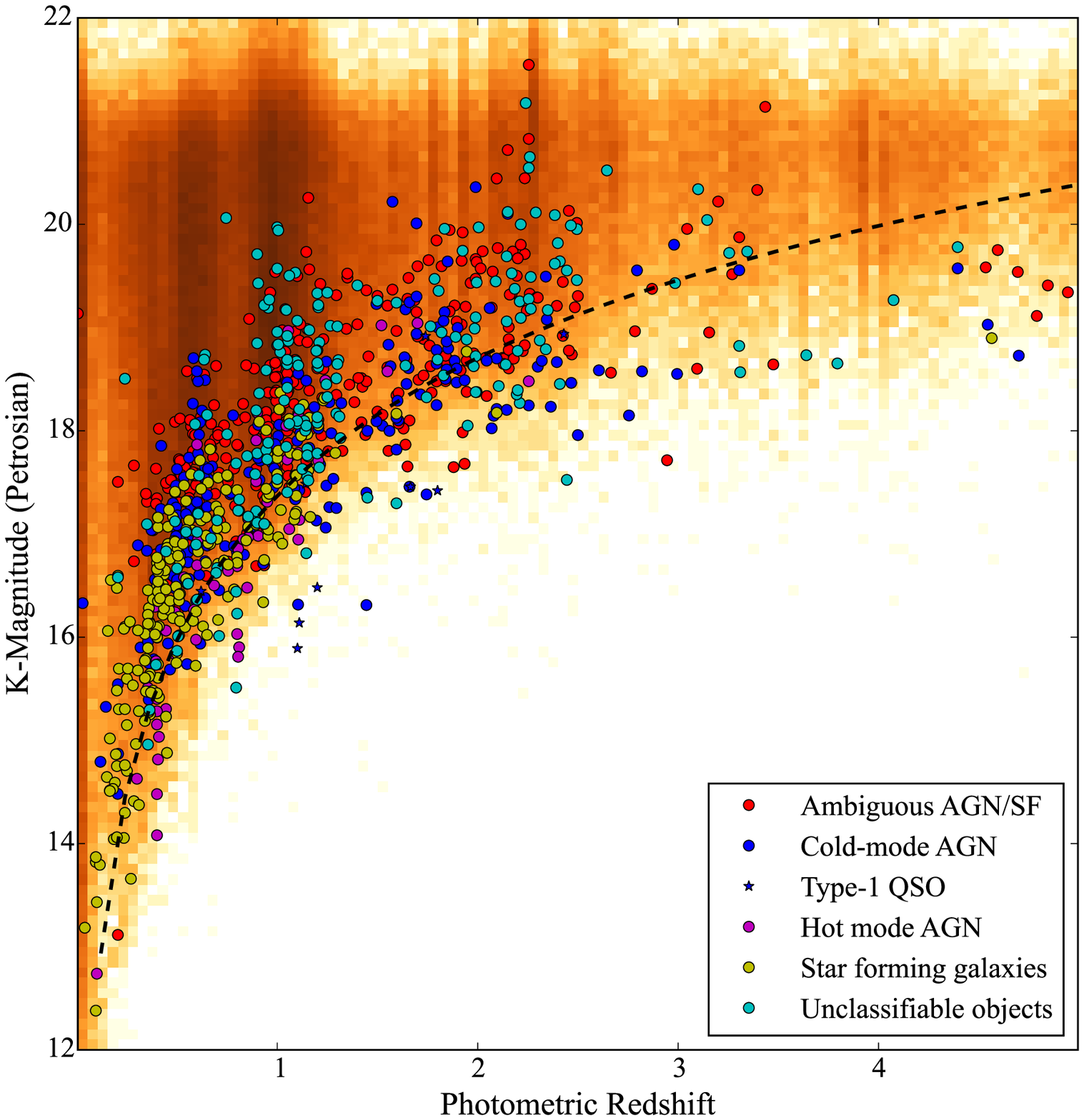}
\caption{The $K-z$ relation by type for the radio sources, with 
the overall $K$-magnitude distribution 
for the SERVS-Lockman data fusion catalog shown as a log-scaled density field. Symbols as in Figure 3. The $K-z$ relation from Willott et al.\ (2003) is shown as the dashed line.}
\label{fig:kzall}
\end{figure}

Figure \ref{fig:kzall} shows the Petrosian magnitudes \footnote{Petrosian magnitudes include the light out to a radius where the surface brightness of the galaxy falls to 0.2 of its mean value within that aperture.}  of the radio sources 
overlaid on the $K-z$ distribution from the entirety of the SERVS Fusion catalog, 
including optical and infrared 
sources which were not matched to radio sources. Despite 
the offset from the $K-z$ relation for the most luminous 
radio galaxies that we observe, the radio matched objects are, at any 
given redshift, among the most luminous objects, and are generally much more luminous than 
the objects found only in the optical, infrared or ultraviolet surveys.

\subsection{Stellar mass}

\begin{figure*}
\plotone{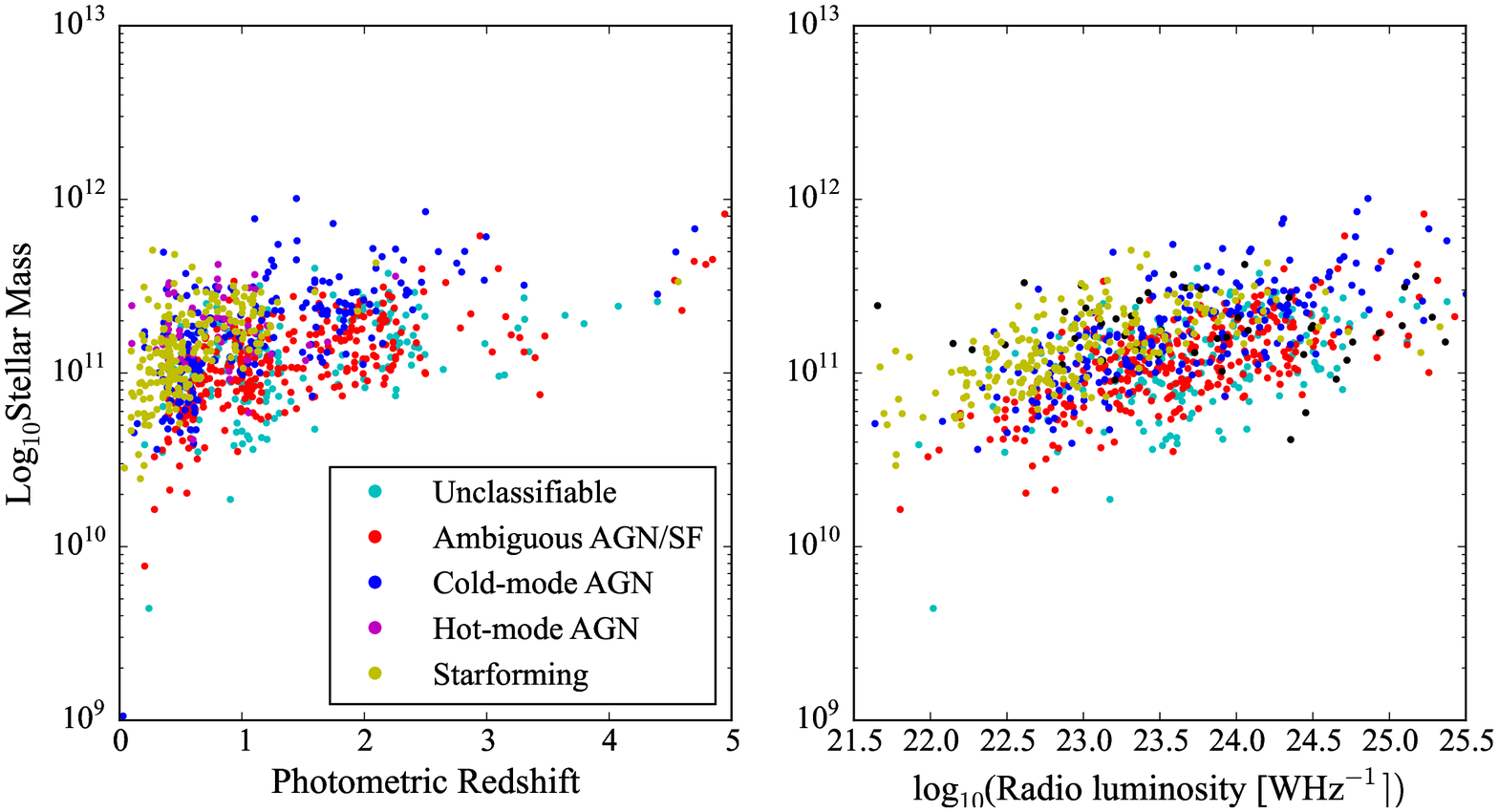}
\caption{Stellar masses of the radio sources as a function of (left) redshift and
(right) radio luminosity.}
\label{fig:masses}
\end{figure*}

We estimated the approximate stellar masses of the Lockman Hole radio sources using the 
3.6~$\mu$m flux density and a 1.4~Gyr stellar population template from 
Bruzual \& Charlot (2003) to calculate the rest-frame $K$-band luminosity, $L_K$. The 3.6$\mu$m wavelength is a good compromise between being close to rest-frame $K$-band at
$z\sim 1$, while at the same time not usually containing significant contamination from hot 
dust emission from the AGN (Caputi 2013). The choice of stellar population template makes only 
a small difference to the $k$-corrections in the near-infrared. We 
then used $K$-band mass-to-light ratio estimates at $z=0-1$ from Drory et al.\
 (2004) and at $z\sim 2$ from Borys et al.\ (2005) to parameterize the 
evolution of the $K$-band mass-to-light ratio in a typical massive 
galaxy as $M/L_K \approx 1.0/(1+1.5z)$. The result is shown in Figure \ref{fig:masses}. 
Some of the largest indicated masses for the cold-mode accretion AGN are 
probably spurious, produced by contamination of the 3.6$\mu$m flux density by 
AGN hot dust emission. For the remainder, we see a fairly well-defined cutoff
at $\sim 10^{11.5}\; M{_{\odot}}$.

We further estimated the stellar masses of the optical and infrared sources which were not 
matched to radio sources, plotting the distribution of stellar masses of the radio-loud
($L_{1.4{\rm GHz}}> 10^{24}\; {\rm WHz^{-1}}$) and radio-quiet populations  
in Figure \ref{fig:kzall}. This Figure shows
that the radio-loud sources tend to be among the objects 
with the largest stellar masses at any redshift see also Bronzini et al.\ (2013), who find that the radio-loud AGN in their work are also in host galaxies significantly more massive than cold-mode AGN or star forming galaxies, though the distribution of masses of  our ``hot mode'' AGN lacks the tail to low masses (below $\approx 10^{10.5}\; M_{\odot}$ seen in figure 9 of Bronzini et al.. A high mean mass for the radio-loud population is expected, based on the work of Best et al.\ (2005) and Simpson et al.\ (2012).

\begin{figure*}
\plotone{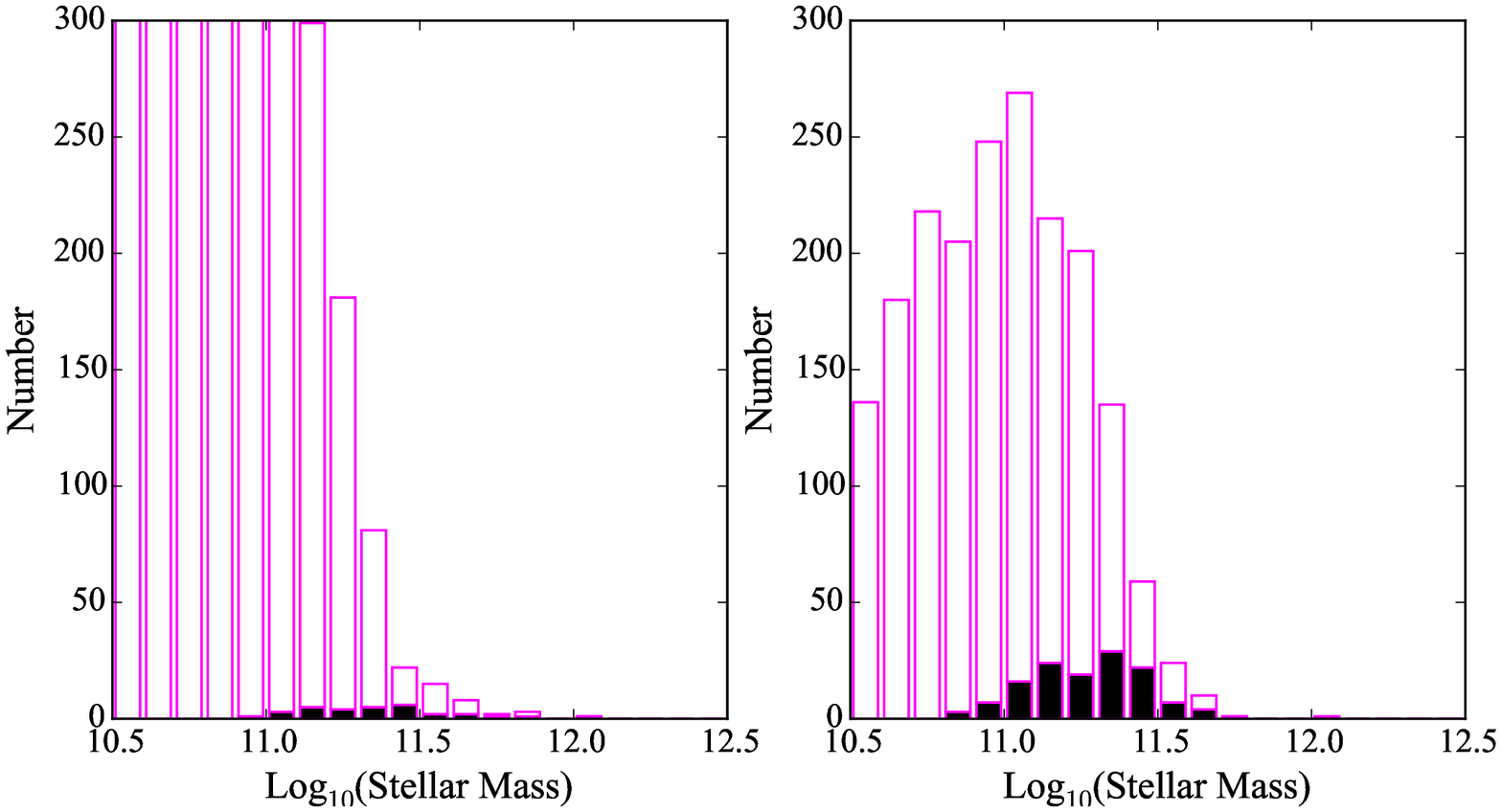}
\caption{Histograms of stellar masses, radio-quiet objects shown in magenta, 
radio-loud objects in black. The left panel is for objects with $0.5<z<1.5$, 
the right panel for objects with $1.5<z<2.5$.}
\label{fig:mhist}
\end{figure*}

Effective AGN feedback depends on a high 
duty cycle of activity (Croton et al.\ 2006), so a high fraction of radio-loud
AGN is needed if radio-jet powered AGN feedback models are to be plausible. Among objects with 
stellar mass greater than 
or equal to $10^{11.5}\; M_{\odot}$ with redshifts $0.5<z<2.5$, we found 19/69, or 28\%, of all 
massive galaxies are radio-loud (Figure \ref{fig:mhist}), consistent with this 
requirement. At $z>2.5$ 
the space density of galaxies with masses $\sim 10^{11.5}M_{\odot}$ is 
very low, $\sim 10^{-6}\; {\rm Mpc^{-3}}$ (Duncan et al.\ 2014), and remains 
similar to the space density of radio sources. It is thus quite plausible
that when a galaxy (and its black hole) reaches a certain critical mass it 
is able to switch on as a radio-loud object with a duty cycle of a few tens of 
percent. Better statistics are needed at 
high-$z$, however, both on the radio source population and on the numbers of 
massive galaxies to understand the evolution of radio source hosts at the 
earliest epochs.

The star-forming galaxies also have relatively high stellar masses,
$\sim 10^{11}\; M_{\odot}$. The likely explanation for this is that these
star-forming galaxies are at the high end of the radio luminosity function
for star-forming objects, with typical radio luminosities
$\sim 10^{22.5}\; {\rm WHz^{-1}}$, forming stars at rates 
$\sim 10 M_{\odot} {\rm yr^{-1}}$ (Chabrier IMF; Karim et al.\ 2011). 
Karim et al.\ (2011) and Zwart et al.\ (2014) 
show that, at the redshifts at which these objects are  mostly
seen in our study ($z\sim 0.3-0.5$), these objects have a typical
specific star formation rate $SSFR \sim 0.1$~Gyr$^{-1}$. Although 
less massive objects tend to have higher $SSFR$, the trend is relatively
weak, so less massive galaxies 
would be expected to have lower absolute 
star formation rates and thus be missing from the survey.

\section{Conclusions}

The results presented in this paper include one of 
the first estimates of the redshift 
distribution and $K-z$ relation for a survey with a large fraction of radio sources $<100\; \mu$Jy. 
The redshift distribution we find features a median value of 
 $z \approx 1$ and a tail out to redshifts of 
$z \approx 5$. This is not entirely consistent with the predicted redshift 
distribution of $\mu$Jy radio sources from the $S^3$ simulations, 
which suggests a larger fraction of sources at $z<0.25$, although the median
redshift is very similar, and issues of over-resolution in the radio may explain the
apparent deficit of very low-$z$ sources.

The galaxies detected in the deep radio survey at $\mu$Jy levels tend to be more luminous 
in $K$-band than objects which were not detected in the radio survey. Nevertheless,
we see evidence that the $K-z$ relation is breaking down, not only due to the increased
fraction of star-forming galaxies, but also because the cold mode 
AGN radio sources 
with 1.4~GHz radio luminosities $L_{1.4}\stackrel{<}{_{\sim}} 10^{24}\;{\rm  WHz^{-1}}$ tend
to exist in less luminous host galaxies. 
Thus lower luminosity radio sources are no longer found 
in only the most massive galaxies, and so, as Simpson et al.\ (2012) point out, 
the $K-z$ relation defined for powerful radio 
sources should not be used for estimating redshifts in 
$\mu$Jy surveys. It is possible that the 
$L_{1.4}\stackrel{<}{_{\sim}} 10^{24}\;{\rm  WHz^{-1}}$ cold-mode AGN have a different mechanism 
for producing radio emission rather than AGN jets. 
Star formation (Kimball et al.\ 2011), or synchrotron
emission powered by shocks in thermal outflows (Zakamska \& Greene 2014) have been suggested as 
possible alternate mechanisms that dominate over jet powered radio
emission at lower luminosities. Hot mode AGN, on the other hand, continue to be found
in only the most massive galaxies at $L_{1.4} < 10^{24}\;{\rm  WHz^{-1}}$ and most likely
continue to be powered by jets. This picture is consistent with only the most massive
galaxies containing black holes that are capable of producing the most powerful radio jets,
and thus with a strong dependence of jet luminosity on black hole mass amongst
AGN (e.g.\ Laor 2000; Lacy et al.\ 2001; Kratzer \& Richards 2014).
We show that the fraction of massive galaxies that are radio-loud continues
to be high (up to $\approx 30$\%) out to $z\sim 2$, a prerequisite for models 
where feedback from radio jets is able to limit star formation 
in their host galaxies, extending the results of Best et al.\ (2005) and
Simpson et al.\ (2012).

\acknowledgements
We thank the referee for a careful reading of the manuscript.
The National Radio Astronomy Observatory is a facility of the National Science 
Foundation operated under cooperative agreement by Associated Universities, Inc.
This work made extensive use of {\sc topcat} (Taylor 2005) for catalog matching
and analysis, and the Virtual Observatory SAMP protocol for communication
between applications. This paper also made use of the SERVS Lockman 
data holdings in the Infrared Science Archive, 
ADS/IRSA.Atlas\#2014/1025/091233\_11822. M.V, acknowledges support from the South African Department of Science and Technology (DST/CON 0134/2014), the European Commission Research Executive Agency (FP7-SPACE-2013-1 GA 607254) and the Italian Ministry for Foreign Affairs and International Cooperation (PGR GA ZA14GR02). J.A. gratefully acknowledges support from the Science and Technology
Foundation (FCT, Portugal) through the research grant PTDC/FIS-AST/2194/2012 and PEst-OE/FIS/UI2751/2014.

\appendix

\section{Accuracy of the photometric redshifts}

\begin{figure}

\plotone{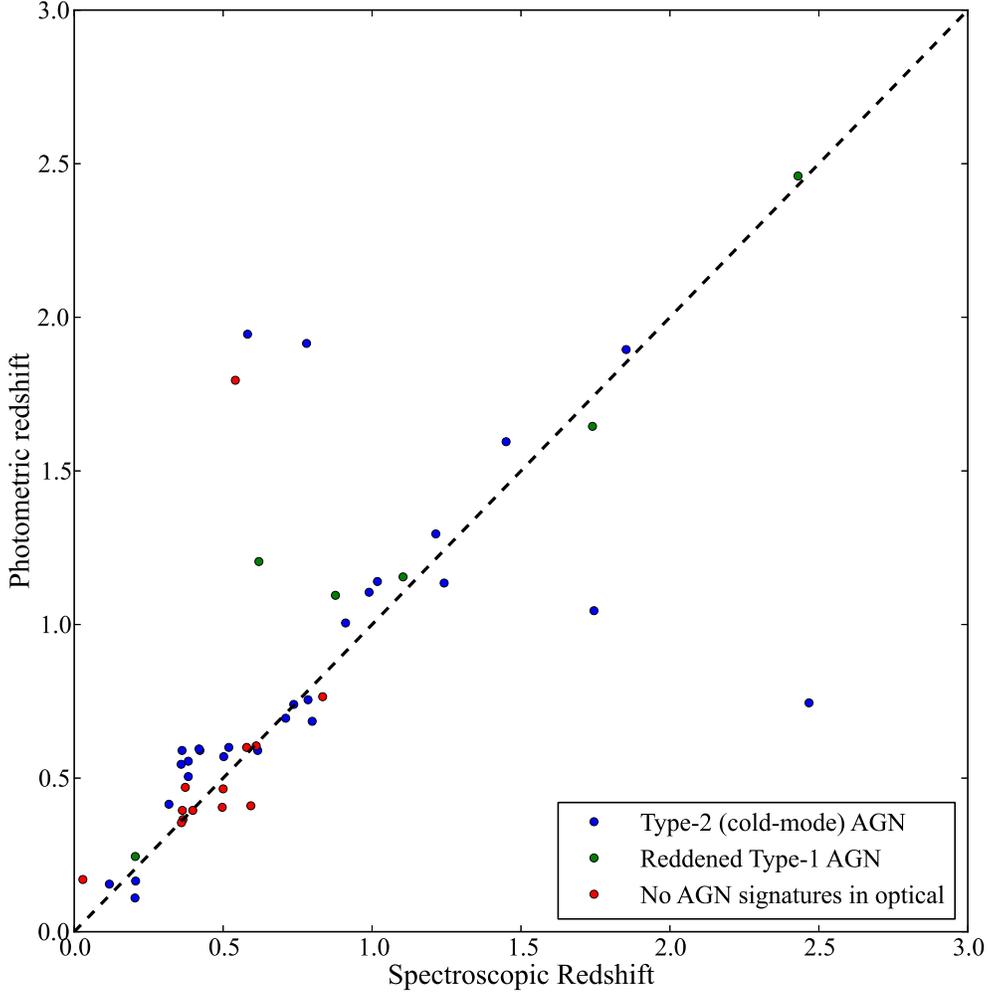}

\caption{Photometric redshift versus spectroscopic redshifts for the objects in the
sample of Lacy et al.\ (2013) with type-2, red type-1 or non-AGN optical/near-infrared spectra and
reliable redshifts}\label{fig:zcomp}
\end{figure}

Spectroscopic redshifts for some of the AGN in this field were obtained by 
Lacy et al.\ (2013). We selected all 53 of the type-2, red type-1 (defined in Lacy et al.\ 2013 as having broad emission lines, but optical colors redder than normal quasars) and ``non-AGN''
objects (i.e.\ AGN candidates selected in the mid-infrared, but which showed no 
AGN signatures in their optical spectra) in
common to the two samples, with spectroscopic redshift quality $q=1$ in Lacy et al.\ (2013).
The plot of spectroscopic versus photometric redshift is shown in Figure \ref{fig:zcomp}.
The two sets of redshifts agree well, though the increasing dominance of the AGN-component
in the higher luminosity AGN above $z\approx 1$ introduces more scatter at high redshifts,
and may also be responsible for some of the catastropic outliers (15\% of the matched objects).
The scatter $\Delta z/(1+z)\approx 0.06$, though there is a small offset for the type-2
objects of $z_{\rm phot} - z_{\rm spec}\approx 0.15$ around $z=0.4$, probably due
to emission line contamination of broad-band magnitudes.
The type-2 objects in this plot are among the most infrared-luminous of the ``cold-mode'' AGN
in this paper, so will be the most prone to contamination by AGN flux in the near-infrared.

\end{document}